\documentclass[twocolumn]{aastex631}

\usepackage{xcolor}
\usepackage{graphicx}
\usepackage{amsmath}
\usepackage{cases}
\usepackage{natbib}

\begin{document}

\title{Kindling the First Stars: I. Dependence of Detectability of the First Stars with JWST on the Pop III Stellar Masses}

\author{Mia Sauda Bovill}
\affiliation{Department of Physics and Astronomy, Texas Christian University, Fort Worth, TX}

\author{Massimo Stiavelli}
\affiliation{Space Telescope Science Institute, 3700 San Martin Drive, Baltimore, MD}

\author{Alessa Ibrahim Wiggins}
\affiliation{Department of Physics and Astronomy, Texas Christian University, Fort Worth, TX}

\author{Massimo Ricotti}
\affiliation{Department of Astronomy, University of Maryland, College Park, College Park, MD}

\author{Michele Trenti}
\affiliation{School of Physics, University of Melbourne, Parkville 3010, VIC, Australia}
\affiliation{ARC Centre of Excellence for All Sky Astrophysics in 3 Dimensions (ASTRO 3D), Australia}

\begin{abstract}

The first Pop III stars formed out of primordial, metal free gas, in minihalos at $z~>~20$, and kickstarted the cosmic processes of reionizaton and enrichment.  While these stars are likely more massive than their enriched counterparts, the current unknowns of their astrophysics include; when the first Pop III stars ignited, how massive they were, and when and how the era of the first stars ended. Investigating these questions requires an exploration of a multi-dimensional parameter space, including the slope of the Pop III stellar initial mass function (IMF) and the strength of the non-ionizing UV background. In this work, we present a novel model which treats both the slope and maximum mass of Pop III stars as truly free parameters while including the physics of the fragmentation of primordial gas. Our results also hint at a non-universal Pop III IMF which is dependent on the efficiency of primordial gas fragmentation. Our relatively simple model reproduces the results from hydrodynamic simulations, but with a computational efficiency which allows us to investigate the observable differences between a wide range of potential Pop III IMFs. In addition, the evolution of the number density of Pop III stars may provide insight into the evolution of the $H_2$ dissociating background.  While the slope of the Pop III IMF does not significantly affect the predicted number density of the first stars, more top heavy IMFs produce Pop III star clusters which are $2~-~3$ magnitudes brighter than their more bottom heavy counterparts. While the Pop III star clusters are too dim for direct detection by JWST, we find they are within the reach of gravitational lensing.

\end{abstract}

\section{Introduction}


The first population of stars which formed  out of material with primordial chemical composition are commonly referred to as Population III stars (Pop III).  In $\Lambda$CDM, the first stars form in minihalos at $z~>~30$ \citep{Abeletal:2002}. Understanding the astrophysical nature of Pop III stars is of fundamental importance, as they begin the process of reionization and the enrichment of the IGM, paving the way for the first Pop II galaxies. With the successful launch of JWST, we stand on the precipice of an unprecedented era in the study of the first stars, however basic questions of the physical properties Pop III stars remain. 

While there is consensus that the typical Pop III star is more massive than their enriched Pop II or Pop I counterparts, the actual mass distribution of Pop III stars remains unknown. Some work suggests masses $>~200$M$_{\odot}$ \citep{Brommetal:1999,Abeletal:2000} with an upper limit of 1000 M$_\odot$ \citep{Ohkuboetal:2009} or higher \citep{Haemmerleetal:2021}.  Other theoretical studies suggest that Population III stars could have a wide range of masses possibly extending down to solar masses and below \citep{Stacyetal:2016,Proleetal:2022,Muhammadetal:2022,Wollenbergetal:2020,Sugimuraetal:2020,Clarketal:2011,Greifetal:2011,Susa:2013,Yoshidaetal:2006,Parketal:2021a, Parketal:2021b}. As such, we do not have an understanding of either the shape of the Pop III stellar initial mass function (IMF) {\it{nor}} the maximum mass for a Pop III star. Both the formation and evolution of these objects are likely different from that of enriched stars in the local Universe and are currently observationally unconstrained. 

One possibility for the study of Pop III stars is to look at ultra metal poor stars in the local universe. If Pop III stars formed at sufficiently low masses, some may have survived to $z~=~0$ \citep{Duttaetal:2020}. However such a detection has not yet occurred and may be complicated by environmental effects such as pollution of the photosphere by supernova ejecta \citep{Sudaetal:2021}. It is unclear whether this implies that low mass ($<1~M_\odot$) Population III stars did not form, have been polluted, or are simply too rare. Alternatively one could attempt to detect the chemical signatures of nucleosynthesis in, and supernova of, Pop III stars in the next generation of stars, however it is unclear whether such signatures would be recognizable \citep{Sarmentoetal:2017,Jeonetal:2021,ChiakiW:2019}.

The second possibility is to detect Pop III stars or their supernova at $z~>~6$. Direct detection of Pop III stars or star clusters is unlikely as the number densities are too low for detection with JWST \citep{Rydbergetal:2013}. Even the tentative identification of the CR7 source with a Pop III object is now mostly discounted \citep{Sobraletal:2015}. However, detection with the aid of gravitational lensing is more promising and has been investigated for magnifications of known lensing clusters \citep{Zackrissonetal:2015,Windhorstetal:2018}. Detection of Pop III stars via gravitational lensing requires detailed lensing models, which exist for a variety of lensing clusters \citep{Lametal:2014,Diegoetal:2016,Diegoetal:2015b,Jauzacetal:2015a,Jauzacetal:2015b}. In addition, individual stars magnified by an order of 10,000 have been detected in caustics \citep{Vanzellaetal:2020,Welchetal:2022} and evidence exists that these high redshift stars are $>~50~M_{\odot}$ \citep{Welchetal:2022}.


A fraction of Pop III stars end their lives as Pair-Instability Supernovae (PISN). These supernovae require high masses ($140~-~260~M_{\odot}$) and have 100 times the energy of their core collapse counterparts. As such, PISN are expected to be extremely luminous and would easily be above the detectability threshold for JWST \citep{Whalenetal:2014,Whalenetal:2013,Moriyaetal:2022}, assuming their rate is high enough \citep{LazarV:2022}.

While both detection via gravitational lensing and PISN are within reach of JWST, determining the astrophysics of the underlying population requires a comparison with theoretical simulations and models of the first stars. Hydrodynamical simulations have now reached parsec resolutions required to study the first stars in a cosmological context \citep{Wiseetal:2012, Xuetal:2016}. However, while some studies have explored the implications for a range of Pop III IMFs \citep{LazarV:2022}, computational expense means most cosmological simulations of the first billion years assume a single power law IMF with an exponential cut off \citep{SkinnerW:2020}. Given the unconstrained nature of the Pop III IMF, a full exploration of parameter space is required for the upcoming observations to inform the astrophysics underlying the formation of the first stars. 

In this work, we introduce a new semi-analytic model and present the new results on the detectability of the first stars with JWST for a range of assumptions for the Pop III IMFs and the maximum Pop III mass. We describe our simulations in Section~\ref{SEC:Sim} and the how the relevant physics is implemented in our model in Section~\ref{SEC:SAM}. Results are given in Section~\ref{SEC.Results}, implications for detection of Pop III stars with JWST are discussed in Section~\ref{SEC.JWST}.

\section{Simulations}
\label{SEC:Sim}
In this section, we describe the numerical simulations and modeling presented in this work. We have generated initial conditions for a set of two high resolution simulations using WMAP9 cosmology \citep{WMAP9} ($\Omega_M~=~0.279,~ \Omega_\Lambda~=~0.721,~h_o~=~0.7$) with MUSIC \citep{HahnA:2011}. The simulations were run with Gadget 2 \citep{Springel:2005} from $z=150$ to $z=6$ and analyzed with AHF \citep{KnollmannK:2009,Gilletal:2004} and Consistent Trees \citep{Behroozietal:2013}. The resolution was chosen to resolve all potential sites of Pop III star formation ($M~>~10^5~M_\odot$) with at least 100 particles.  While test runs were done on the UDF cluster at STScI, both high resolution boxes in Table~\ref{TAB.simulations} were run and analyzed on the University of Maryland HPCC Deepthought2.

\begin{table}[h!]
\centering
\caption{Simulations}
\begin{tabular}{ccc}
\hline
simulation & $L$ & $N$ \\
 & (Mpc/h comoving) &  \\
\hline
2M\_512 & 2 & 512 \\
4M\_1024 & 4 & 1024 \\
\end{tabular}
\label{TAB.simulations}
\end{table}
\label{TAB:Sim}

\subsection{Model}
\label{SEC:SAM}

In this section we describe a novel semi-analytic model which will allow us to investigate the unexplored parameter space of the formation, properties, and fates of Pop III stars. Critical physics incorporated into our model includes; a variable Pop III IMF, limits for the fragmentation of primordial gas, an external and self-consistent LW background, and a model for the enrichment of the Pop III host halos and the nearby IGM by supernova.

\subsubsection{Mass Thresholds for Pop III Star Formation}

The primary driver of whether a given halo will be able to form stars is the strength of the Lyman Werner (LW) background and is determined by four mass thresholds. Here we provide the analytical forms of those thresholds for the WMAP9 cosmology used in this work. Throughout, we assume a halo is pristine if $Z~<~10^{-5}~Z_\odot$. In the absence of a dissociative UV or X-ray background a dark matter halo must simultaneously meet the criteria for $H_2$ cooling and Jeans' collapse to be massive enough for its gas to cool and collapse \citep{Tegmarketal:1997}. This gives us:
\begin{equation}
M_{vir}^{III} = max\begin{cases}
		M_{H_2} = 1.905 \times 10^5 M_\odot~ (\frac{1+z}{31})^{-1.5}  \\
		~\\
   		M_J = 1.36 \times 10^5 M_\odot ~(\frac{1+z}{31})^{-2.071} 
  	\end{cases}
\label{EQ.MIII}
\end{equation}
where $M_{vir}^{III}$ is the minimum halo mass required for Pop III star formation at a given redshift without a dissociative background, $z$. Note, $M_{vir}^{III}$ is determined by the Jeans' collapse, $M_J$  at lower redshift ($z~<~16.2$) and by the threshold for $H_2$ cooling, $M_{H_2}$ at higher redshift, ($z~>~16.2$).

Before discussing the mass threshold for the LW background, we note that regardless of the strength of the LW background, a halo above the HI cooling limit, $M_{HI}$, will be able to cool its pristine gas via HI pathway:
\begin{equation} 
M_{HI}~=~7.75\times10^6~M_\odot ~\bigg(\frac{1+z}{31}\bigg)^{1.5}
\end{equation}
\label{EQ.MHI}

where $M_{HI}$ is the HI cooling mass at a given redshift, $z$. In addition to being able to cool via neutral hydrogen, halos with $M>M_{HI}$ will be able to self shield $H_2$, even in the presence of the LW background.

\subsubsection{Lyman Werner Background}

Before Pop III stars explode, enriching their host halos and surrounding IGM with metals, they emit non-ionizing UV radiation which builds up to form the LW background. As LW photons have a mean free path of approximately $150^{th}$ of the cosmic horizon \citep{Ricotti:2016,Ricottietal:2000}, corresponding to $\sim 4$ Mpc {\it{physical}} at $z~\sim~10$ss \citep{Stiavelli:2009,GloverB:2001}, more than ten times the size of our simulation box, we include an externally generated LW background in our model. 

For a given LW intensity, $J_{21}$, the minimum mass required for a halo to form Pop III stars is given by:
\begin{equation}
M_{LW} = 6.44 \times 10^6 M_\odot  ~J_{21}^{0.457} ~\left(\frac{1+z}{31}\right)^{-3.557}
\label{EQ.LW}
\end{equation}
where $M_{LW}$ is the mass threshold required for Pop III star formation in the presence of a LW background with intensity $J_{21}$ at a redshift, $z$. 

\begin{figure}
\centering
\includegraphics[width=\columnwidth]{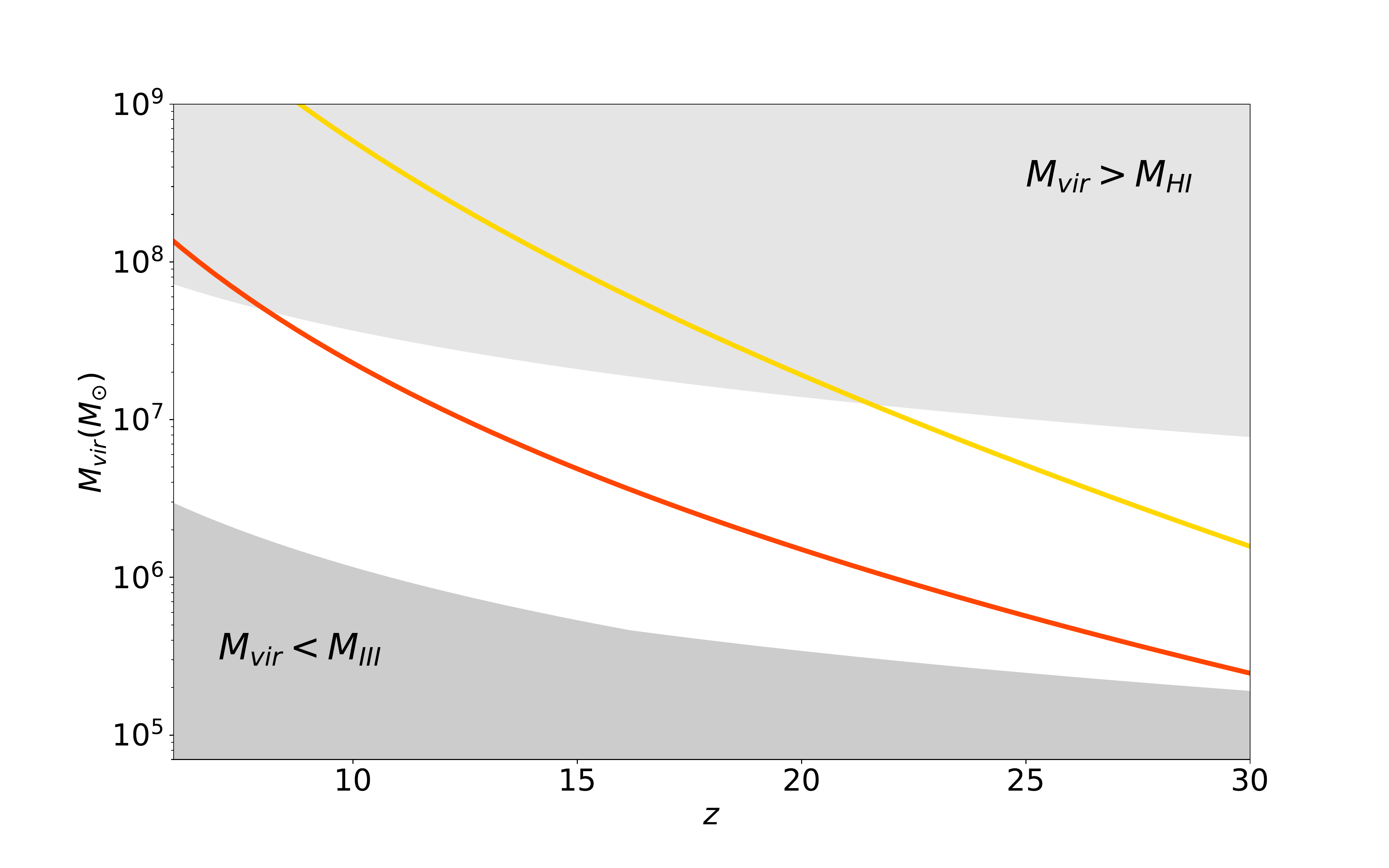}
\caption{Halo viral mass versus redshift for various thresholds for Pop III star formation. The dark grey area on the bottom shows the region where a halo will not be able to form Pop III stars because there is either not enough gas to become Jeans unstable or enough $H_2$ to cool the gas (Equation~\ref{EQ.MIII}). The pale gray shaded area at the top is where Pop III stars will always be able to form since $M_{vir} > M_{HI}$ and there is sufficient gas to self shield the $H_2$ from the LW background. The color lines show the mass thresholds for Pop III star formation for a strong (gold) and weaker (red) LW background from TS09 (Equation~\ref{EQ.LW}).}
\label{FIG.TS09}
\end{figure}

The $M_{LW}(z)$ dependence for a range of $J_{21}$ values are taken from a from a subset of the $J_{21}(z)$ models from TS09 (Figure~\ref{FIG.TS09}). We select the TS09 model by sorting the full range of models into three groups based on the strength of the $J_{21}(z)$. Of the seven cases with a LW background, four (standard Pop III model with a Sheth Tormen  or a Press-Scheter halo mass function and multiple Pop III stars per halo with $\epsilon_{PopIII} = 0.005$) have a medium LW background.  A reduced $f_{esc}$ produces the one low LW background case, and the two high cases have a strong external $J_{21}$ field and multiple Pop III stars forming with $\epsilon = 0.05$, respectively.

In this work, we consider the low and high LW background cases (Figure~\ref{FIG.TS09}). While the modeling of the Pop III population in TS09 is different from our model, the low LW case is roughly consistent with the LW background generated by the stellar populations in our simulations. We have included the strong LW case to investigate how Pop III star formation would change near an over-dense region capable of generating such a background.

Note, $J_{21} = J_{21}^{external} + J_{21}^{consistent}$, including both the external LW backgrounds from \cite{TrentiS:2009} ($J_{21}^{external}$) and a self-consistent component generated by all Pop III stars and Pop II stars $>~8~M_\odot$ ($J_{21}^{consistent}$). For details of the $J_{21}^{consistent}$ calculation, see \cite{TrentiS:2009}.

\subsubsection{Pop III IMF}

\begin{figure}
\centering
\includegraphics[width=\columnwidth]{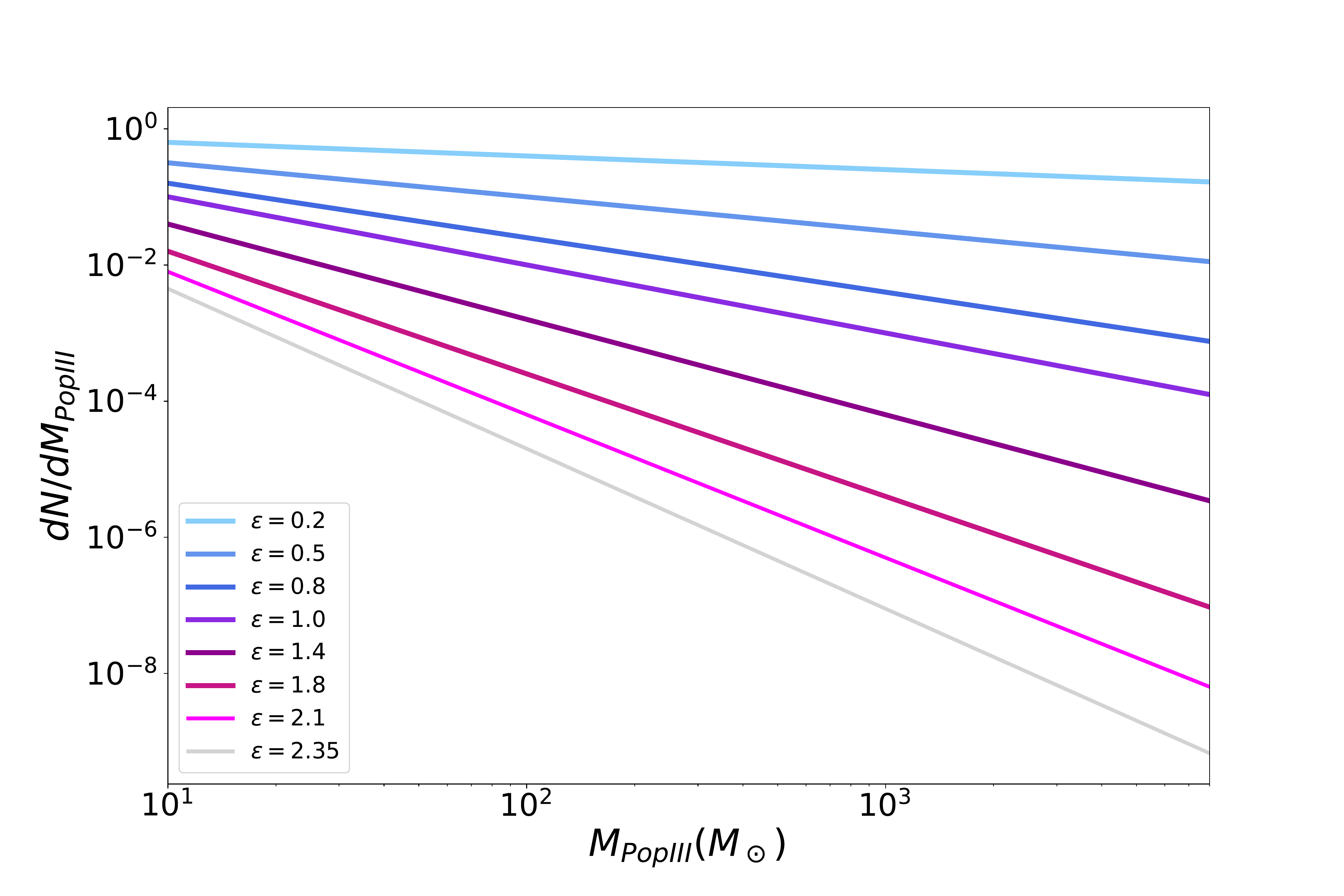}
\caption{The range of slopes for the Pop III IMF considered in this work. The color coding in this figure will be used throughout this work for the various IMF slopes. }
\label{FIG.III_IMF}
\end{figure}

In this work, we assume the Pop III IMF is described by a power law, However we consider the slope of the power law to be a free parameter. For the IMF slope we explore $\alpha~=~[0.2, 0.5, 0.8, 1.0, 1.2, 1.5, 1.8, 2.1, 2.35]$ (Figure~\ref{FIG.III_IMF}) where:
\begin{equation}
dm/dN~=~Am^{-\alpha}
\end{equation}
where $m$ is the mass of a Pop III star, $\alpha$ is the assumed slope of the Pop III IMF, and $A$ is the normalization for the mass of gas available in the halo.

For all IMF slopes we assume a minimum Pop III mass of $10~M_\odot$ and a maximum mass of $1000~M_\odot$. In any given halo, the number of Pop III stars which can be formed is set by the maximum mass of Pop III stars, $M_{III}^{max}$, and the physics of the fragmentation of primordial gas. The former is given by $M_{III}^{max}~=~\epsilon_{III}^{max}f_bM_{vir}$, where $M_{vir}$ is the virial mass of the halo, $f_b$ is the baryon fraction and $\epsilon_{III}^{max}$ is the maximum possible Pop III star formation efficiency, a parameter provided for the model. Our fiducial value is $\epsilon_{III}^{max}~=~0.01$.

The latter quantity is calculated with the assumption that the lack of an efficient coolant means primordial gas will not fragment as effectively as enriched gas. An estimate for the maximum number of fragments halo with $M_{vir}$ at a given redshift, $z$ can be derived from Jeans collapse. In order for gas to cool, collapse and form stars, the total gas mass, $M_g > M_J$. We assume the initial temperature of the gas is $T_{vir}$ and the gas cools to a final temperatures, $T_f$. If we assume all of the gas in the halo collapses into a single fragment, we recover the Jeans mass requirement for Pop III star formation given in Equation~\ref{EQ.MIII}, however, this is only true if the gravitational collapse timescale, $\tau_{coll}$, is less than the cooling time scale, $\tau_{cool}$. For massive enough halos $\tau_{coll} > \tau_{cool}$, which means the gas cools faster than it collapses, potentially resulting in additional fragmentation \citep{Stiavelli:2009}.

To determine the maximum amount of gas fragmentation in a pristine halo with a given $M_{vir}$ at a given redshift, $z$, we start with the simple idea that 
\begin{equation}
NM_J \le M_g = f_b M_{vir}
\label{EQ:gasmass}
\end{equation}
where $N$ is the maximum number of fragments, $M_g$ is the available gas mass, $f_b$ is the cosmic baryon fraction and 
\begin{equation}
M_J = 8.4 \times 10^6 \bigg(\frac{M_{vir}}{10^6 M_\odot}\bigg)\bigg(\frac{T_f}{T_o}\bigg)^2
\label{EQ.Jeans}
\end{equation}
where $M_J$ is the Jeans mass, $M_{vir}$ is the virial mass of a halo, $T_o~\approx~T_{vir}$ is the initial temperature of the gas and $T_f$ is the final temperature of the gas.

\begin{figure}
\centering
\includegraphics[width=\columnwidth]{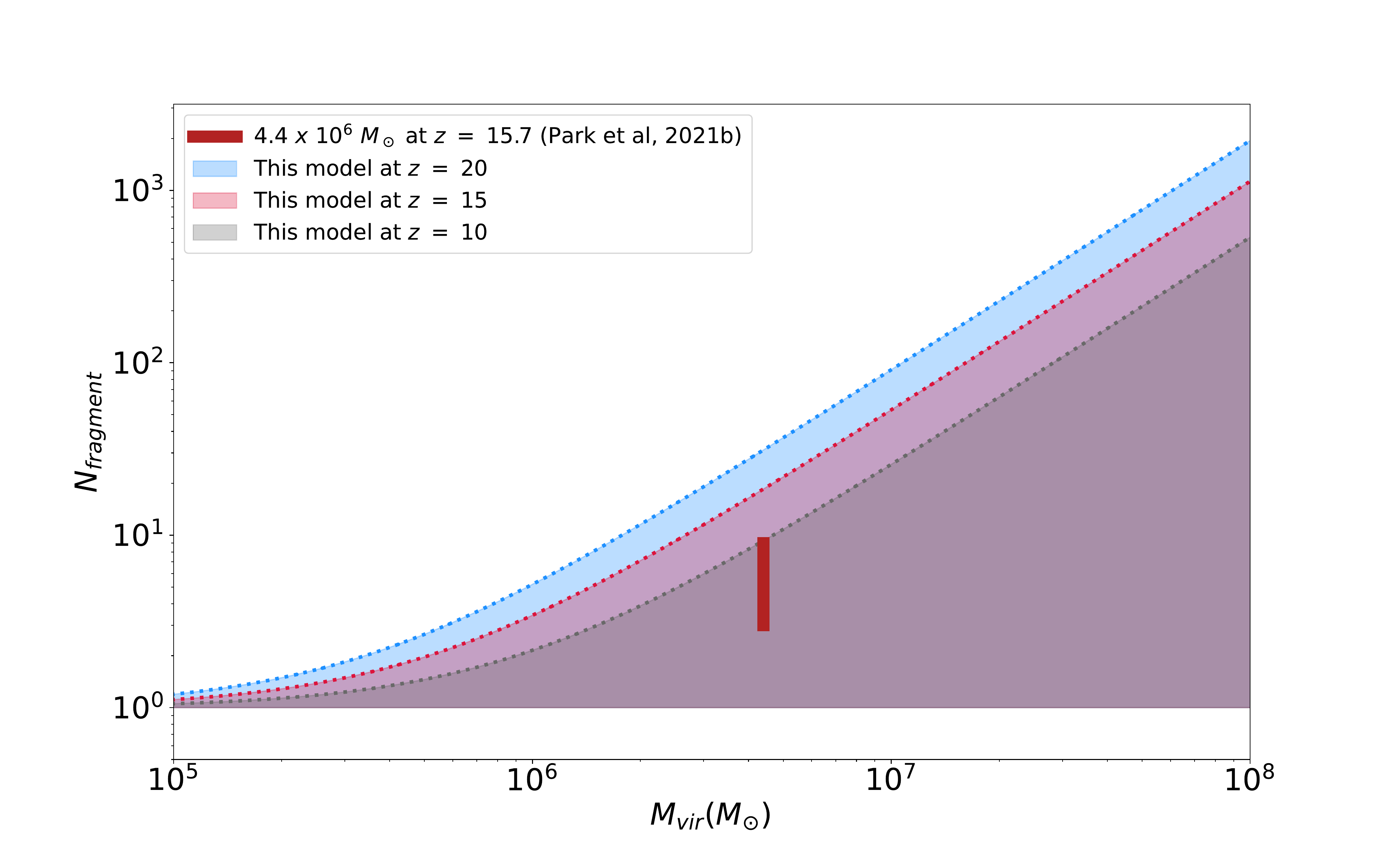}
\caption{The shaded regions show the number of fragments in a pristine halo as a function of virial mass. The upper edge of the shaded regions is given by Equation~\ref{EQ.Nfrag} for $z~=~20$ (blue), $z~=~15$ (red), and $z~=~10$ (dark grey), and the lower bound by $N_{frag}~=~1$ where $M_J~=~f_bM_{vir}$. The dark red bar shows the range of fragmentation for a $4.4~\times~10^6~M_\odot$ halo at $z~=~15.7$ from \cite{Parketal:2021b}.}
\label{FIG.Nfrag}
\end{figure}

Assuming the fraction of baryons in the halos is equal to the cosmic baryon fraction, for the gas to become Jeans unstable ($M_J < M_b$), the gas must cool until $T_{vir}/T \ge 6.97$. When this is combined with Equation~\ref{EQ:gasmass} and Equation~\ref{EQ.Jeans} this gives us
\begin{equation}
\bigg(\frac{T_{o}}{T_f}\bigg)^2 \ge \frac{f_b}{8.4}N = 6.97^2N
\end{equation}
. Given that $T_o = T_{vir}$ and $T_f~\approx~120K$, the minimum temperature to which $H_2$ can cool gas \citep{Abeletal:2000} we have:
\begin{equation}
N_{frag} \le 9.12\bigg(\frac{M_{vir}}{10^6 M_\odot}\bigg)^{4/3} \bigg(\frac{1+z}{31}\bigg)^2
\label{EQ.Nfrag}
\end{equation}
where $N_{frag}$ is the maximum fragmentation of primordial gas in a halo of $M_{vir}$ at a redshift $z$.  The expected number of fragments for a pristine halo in our model for a given mass at three redshifts is shown in Figure~\ref{FIG.Nfrag}. We find our model is consistent with the amount of fragmentation seen in high resolution hydrodynamical simulations \citep{Parketal:2021b}. 

Physically this means a halo of $M_{vir}$ at a redshift $z$ will have primordial gas fragmentation into $<~N_{frag}$ fragments. However, multiple Pop III stars may form per fragment, as high resolution hydrodynamic simulations suggest as many as six Pop III stars can form per fragment \citep{Susaetal:2014}. In our model, this is parametrized as a constraint on the maximum number of Pop III stars $N_{III}$ where $N_{frag}~<~N_{III}~>~6N_{frag}$.\\

We find that for a given $\epsilon_{III}^{max}$, $N_{frag}$ decreases for all Pop III IMFs and the effect is larger for the more top heavy Pop III IMFs. In addition, we find that as average number of stars per fragment increases, $N_{frag}$ decreases regardless of the choice of $\epsilon_{III}^{max}$ and the Pop III IMF, however, the effect is greater for more top heavy Pop III IMFs and lower $\epsilon_{III}^{max}$.

The explanation of both trends is in the combination of two effects. First, it is important to note that Equation~\ref{EQ.Nfrag} gives a {\it{maximum}} number of fragments for a primordial halo with $M_{vir}$ at a given redshift. Our model allows the primordial gas to fragment less than this limit. Second, that the total mass of Pop III stars in any halo, $M_{III}$, in limited to $M_{III}~\le~\epsilon_{III}^{max}f_bM_{vir}$.
In this work, a Pop III star forming halo must meet both criteria. Therefore, lower values $\epsilon_{III}^{max}$ limit the total mass budget available for Pop III stars, and more top heavy Pop III IMFs place more of that mass budget into fewer more massive stars. The more stars which form in a typical fragment, the less gas fragmentation is required  before the mass budget for Pop III stars is exceeded.

For the power law IMF we randomly sample the power law with a $m_{min} = 10 M_\odot$. The power law is resampled until a distribution of masses is produced for which $M_{tot} \le M_{PopIII}$ and $N_{tot} \le N_{III}N_{frag}$. For the Pop III IMF power law slopes, when both the mass of the gas reservoir and the fragmentation of the primordial gas are taken into account, our results become insensitive to the choice of a Pop III star formation efficiency. We define the de facto Pop III star formation efficiency as $M_{PopIII}/f_b M_{vir}$. 

\begin{figure*}[ht]
\centering
\includegraphics[width=\columnwidth]{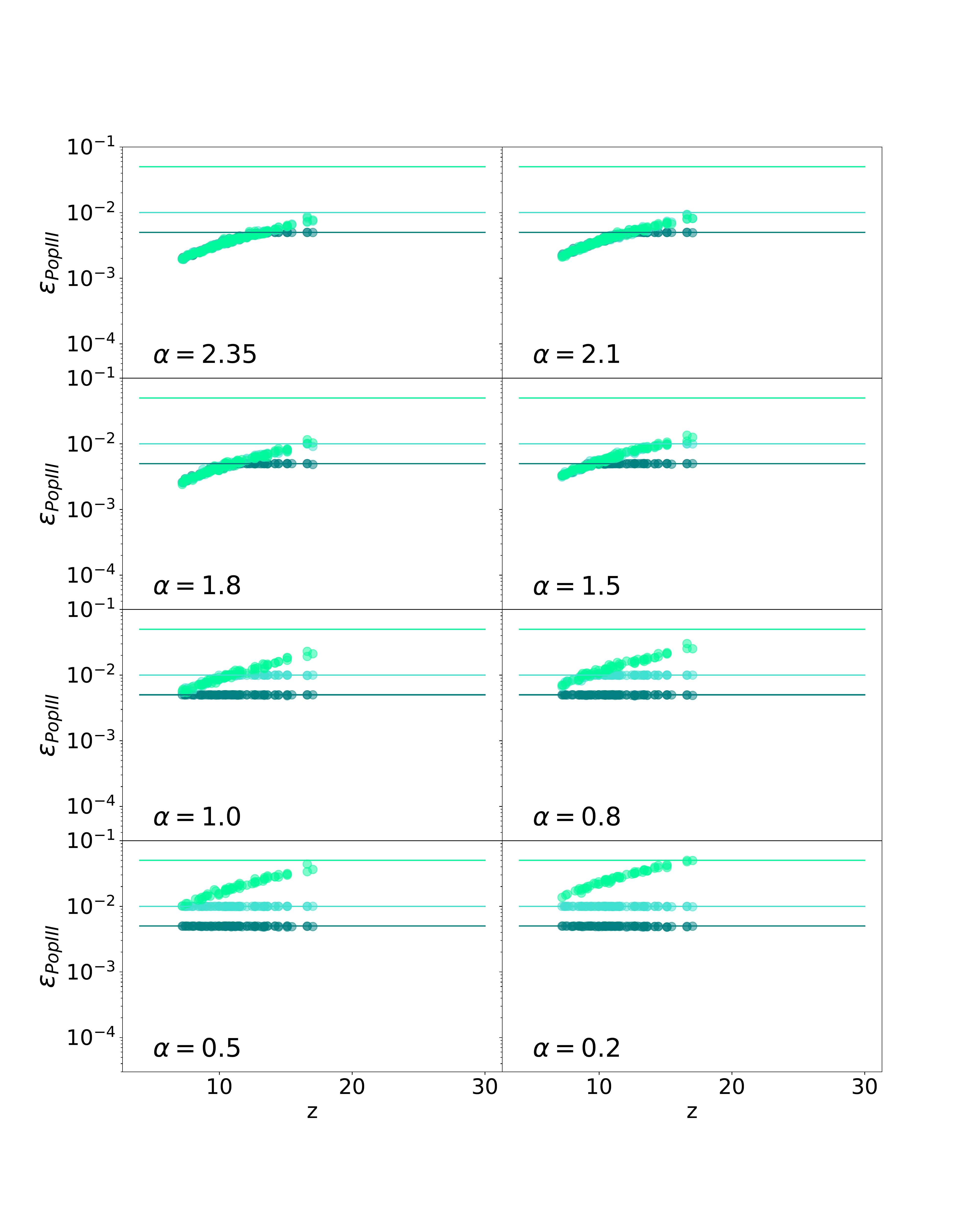}
\includegraphics[width=\columnwidth]{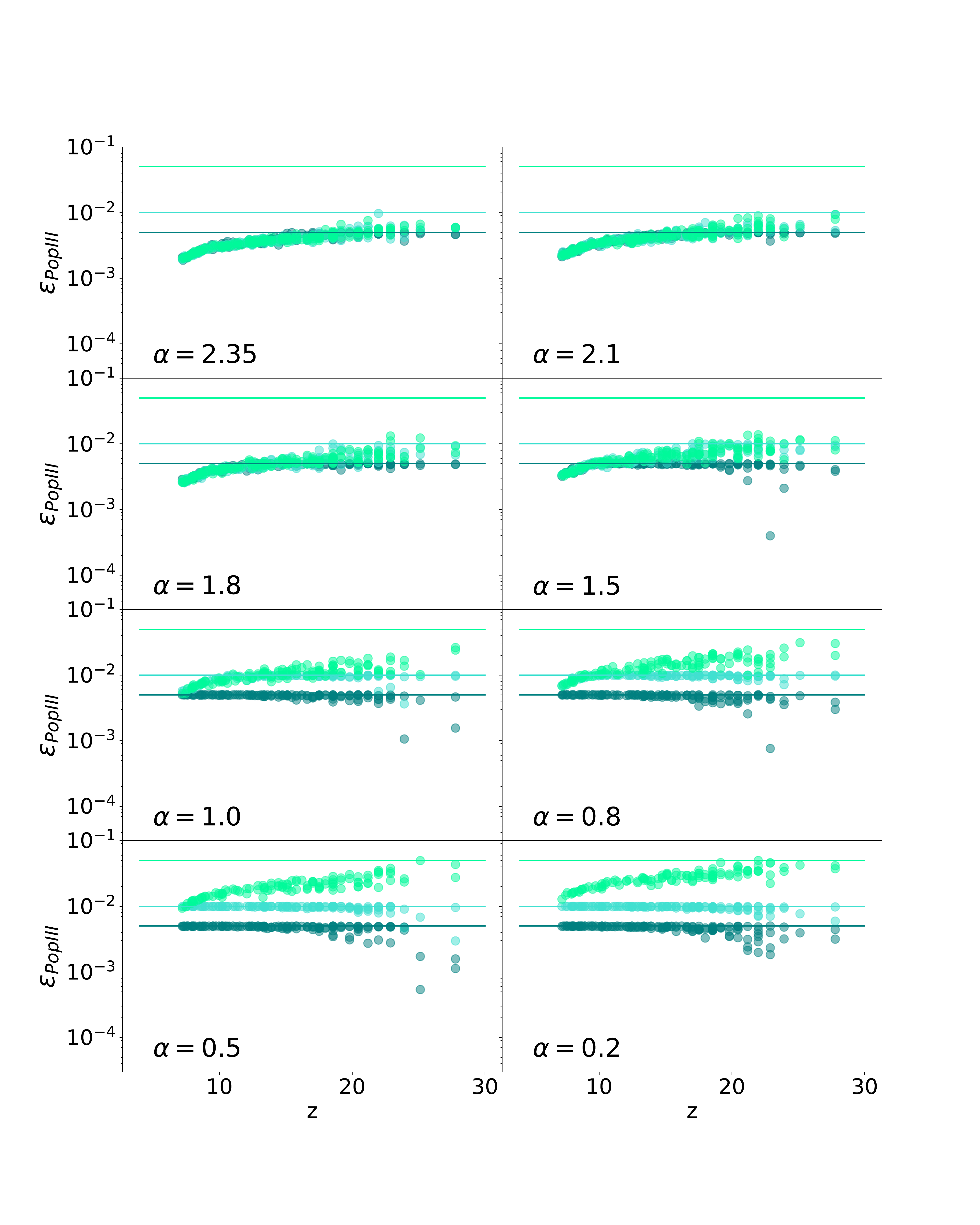}
\caption{The actual efficiency of Pop III star formation defined as $\epsilon_{PopIII}~=~M_{III}/f_b M_{vir}$ for our range of Pop III IMF slopes (Figure~\ref{FIG.III_IMF}) and the low (right) and high (left) external LW backgrounds (Figure~\ref{FIG.TS09}) for three values of $\epsilon_{III}~=~[0.005$(dark blue)$,~0.01$(turquoise)$,~0.05$(green)$]$ inputted into the model. Note that for the bottom heavy IMFs, $\epsilon_{PopIII}$  are independent of the value of $\epsilon_{III}$ inputted into the model.}
\label{FIG.SFE}
\end{figure*}

As seen in Figure~\ref{FIG.SFE}, the de facto Pop III star formation efficiencies from our relatively simple model are in reasonable agreement with Pop III star formation efficiencies measured in high resolution hydrodynamic simulations of the first stars \citep{SkinnerW:2020}.

Figure~\ref{FIG.SFE} shows that for bottom heavy IMFs, the measured Pop III star formation efficiency is independent of our choice of $\epsilon_{III}$. This is due to the limited fragmentation of the primordial gas, which places an upper limit on the number of Pop III stars which can form. For the more top heavy IMFs, our choice of $\epsilon_{III}$ does affect the defacto star formation efficiency. At $z~>~20$, there is increasing scatter in the defacto Pop III star formation efficiency with an increasingly top heavy IMF.  This is likely driven by the stochastic population of the high mass end of the IMF.  This suggests that the efficiency of the fragmentation of primordial gas is a significant driver of the Pop III IMF. The more efficient the fragmentation, the more bottom heavy the IMF. The fragmentation of primordial gas can be increased by turbulence and torques from nearby halos. As both of these effects increase with decreasing redshift, the changing fragmentation of primordial gas with redshift may result in a non-universal Pop III IMF which evolves with both halo mass and redshift.

\subsubsection{Metal Enrichment}
\label{SEC.metal}

To determine when the transition from unenriched to enriched star formation occurred, we must determine when and where there has been metal enrichment. The metal enrichment due to the first stars  is driven by the fates of the first stars.  In our model this is given by: 
\begin{equation}
	\begin{cases}
		m < 140 M_{\odot}  & CCSN\\
   		140 M_\odot < m < 260 M_\odot & PISN\\
		260 M_\odot < m & DCBH
  	\end{cases}
\end{equation}
from \cite{Hegeretal:2003}. For Pop III stars which die as DCBH, we assume the host halo has been enriched, but that any metal enrichment remained trapped in the host halo and did not enrich the surrounding IGM. For enrichment by Pop III stars which die by CCSN and PISN we use the prescription discussed below.

The Pop III era is ended by enrichment of halos and the IGM by the first supernova. In order for a halo to form Pop III stars its gas must have $Z~<~Z_{crit}$. We assume a $Z_{crit}~\approx~-5$, however our results are not sensitive to this choice.

For a halo to be considered pristine ($Z~<~Z_{crit}$) in our model it must meet two criteria; no progenitor halo has formed either Pop III or Pop II stars and no progenitor halo was polluted by SN ejecta from a neighboring halo. Determining the latter criteria requires a simple model to determine whether the SN ejecta have enough mechanical energy, $E_{SN}$ to leave the halo and how far into the IGM the ejecta spread if they do. 
Supernova ejecta are contained within the host halo if:
\begin{equation}
M_{contain} = 7.96 \times 10^6 M_\odot \bigg(\frac{E_{SN}}{10^{51} erg}\bigg)^{3/5}\bigg(\frac{1+z}{31}\bigg)^{-3/5}
\end{equation}
where $E_{SN}$ is the amount of kinetic energy energy generated by Pop III supernovae, $f_{kin}(10^{51} ergs N_{CCSN} + 10^{53} ergs N_{PISN}),$ in a halo with $M_{vir}$ at a redshift $z$. In our fiducial model we assume $f_{kin}~=~30\%$ of the SN energy is converted into mechanical energy. In our model we assume all supernova from a given burst of star formation have, at roughly the same time exploded and ejecta have dispersed within the $10$~Myr before the next simulation snapshot. If $M_{vir} \ge M_{contain}$ then we assume the enriched supernova ejecta remain confined to the halo and do not pollute the surrounding IGM, while if $M_{vir} < M_{contain}$ then a fraction of the enriched supernova ejecta escapes from the parent halo and pollutes the surrounding IGM.

In the latter case we can use physical arguments to estimate the radius of the enriched ejecta $R_{ej}$. Outside of the initial halo, we assume a constant IGM density, $\rho_{IGM}$ where $\rho_{IGM}$ is simply the mass density of non-virialized mass at a redshift, $z$. Given that the physical IGM in the immediate vicinity of a halo is clumpier and denser than these assumptions, the $R_{ej}$ we calculate can be seen as an upper bound on the results of a more complex treatment. 

For the mass enclosed inside the ejecta radius we assume the enriched gas expands until the thermal speed of the gas is equal to the escape speed from the halo.
\begin{equation}
v_{esc}^2  = v_{th}^2 = \frac{2 E_{SN} }{3 f_b M(<R_{ej})}
\label{EQ:vesc}
\end{equation}
where $E_{SN}$ is the mechanical energy of supernova from a halo, $f_b$ is the baryon fraction, $M_(<R_{ej})$ is the mass enclosed within the host halo and a smooth IGM with $r<R_{ej}$, and $v_{esc}$ is the escape velocity at $R_{ej}$ and $v_{th}$ is the thermal speed given by $E_{SN}$. This gives us:
\begin{equation}
\frac{E_{SN}}{3 f_b G} = \frac{M^2(<R_{ej})}{R_{ej}}
\end{equation}
For our assumption of a smooth IGM we can write $M(<R_{ej}) = M_{vir} + \frac{4\pi}{3}\rho_{IGM}(R_{ej}^3 - R_{vir}^3)$ where $\rho_{IGM} = (1 - f_{coll}(z))\rho_o (1+z)^3$ and $M_{vir} = \frac{4\pi}{3}\xi_{vir}R_{vir}^3$ and $\xi_{vir} \sim 178 \rho_o (1+z)^3$. Combining these with Equation~\ref{EQ:vesc} gives:
\begin{equation}
\frac{E_{SN}}{f_b G} = \frac{(M_{vir} + \frac{4 \pi}{3}\rho_{IGM}(z)(R_{ejecta}^3 - R_{vir}^3))^2}{R_{ej}}
\end{equation}
where $E_{SN}$ is the energy of a supernova in a halo of $M_{vir}$ and $R_{vir}$ at a redshift $z$ expanding into an a smooth IGM.

\section{Results}
\label{SEC.Results}

\subsection{Lyman Warner Background}

\begin{figure*}
\includegraphics[width=\columnwidth]{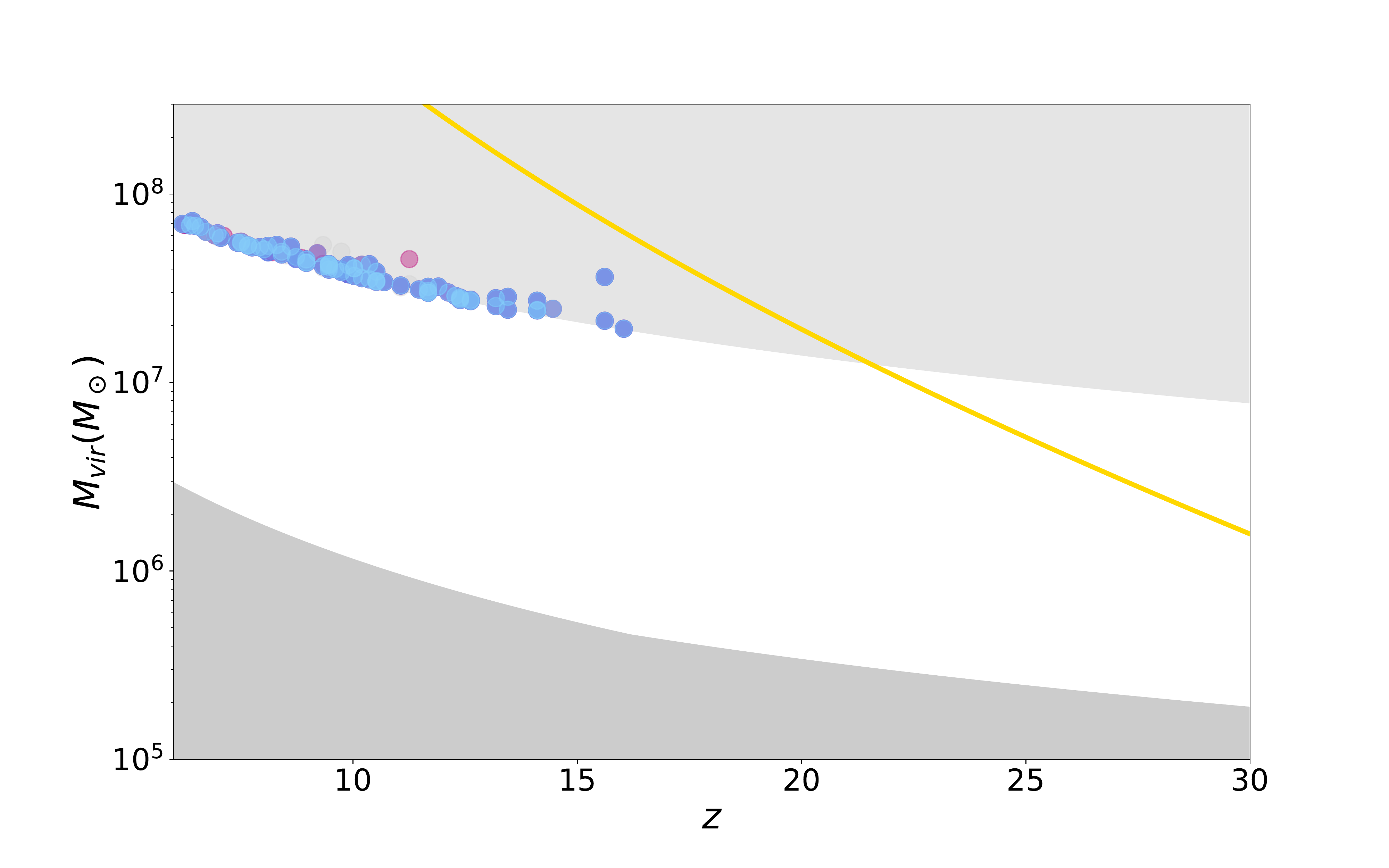}
\includegraphics[width=\columnwidth]{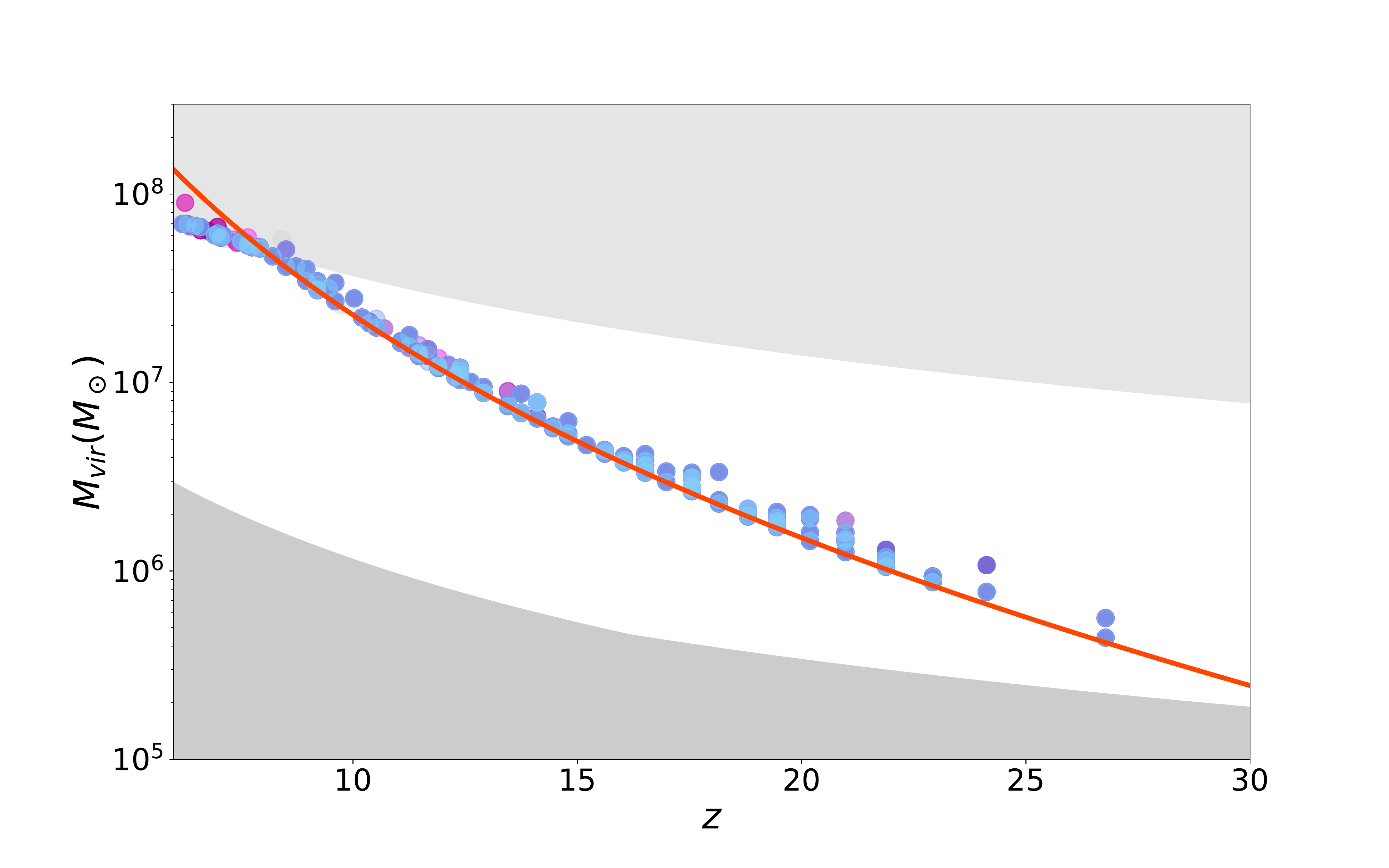}
\caption{The virial mass versus redshift for the halos hosting Pop III stars for a strong (left) and weak (right) LW backgrounds. The dark grey shaded region is $M_{vir} < M_{III}(z)$, where no Pop III stars will form due to insufficient mass or coolant. The light grey shaded region is where $M_{vir}~>~M_{HI}$, where HI cooling and self shielding of $H_2$ allow Pop III stars to form regardless of the strength of the LW background. Mass thresholds due to the weak (red) and strong (gold) LW backgrounds are shown by the dotted lines. Each point is a halo and the colors correspond to the Pop III IMFs in Figure~\ref{FIG.III_IMF}.}
\label{FIG.Mvir_z_III}
\end{figure*}

We first look at when and in what halos Pop III stars formed. Figure~\ref{FIG.Mvir_z_III} shows the virial mass versus redshift of the dark matter halos which formed Pop III stars for the full range of the Pop III IMF slopes and LW backgrounds considered in this work. As expected, the redshift at which Pop III stars form decreases as the strength of the LW background increases. Concurrently, the typical mass of halos forming Pop III stars increases as the strength of the LW background increases. This is due to the increased suppression of star formation in low mass halos. For detectability, the best option is Pop III stars forming at $z~<~15$ in halos above the atomic cooling limit. As seen in Figure~\ref{FIG.Mvir_z_III}, these atomic cooling Pop III halos are present for all external LW backgrounds and pop III IMFs explored in this work.

While variations in the number density of Pop III stars forming halos is secondary to the differences driven by the strength of the LW background, we do observe trends. We note that as the Pop III IMF becomes more top heavy, the number density of Pop III star forming halos decreases, especially at $z~<~12$. This can be explained by the more efficient enrichment of the IGM due to the higher number of energetic PISN supernova for the more top heavy IMFs. In this work, the end of the Pop III era is driven by the external enrichment of low mass halos and the IGM by enriched ejecta from Pop II star forming halos. We discuss this in more detail in \S~\ref{SEC.IMF}.

We now look at the case of a strong LW background in more detail. The strong LW background used here is that generated by the \cite{TrentiS:2009} model with massive Pop III stars forming with $\epsilon = 0.05$. This stronger background approximates the case of Pop III star formation near an over-dense region or proto-cluster. Effectively, this strong LW background suppresses all Pop III star formation in halos with $M_{vir}~<~M_{HI}$ in the following manner.  Via dissociation of $H_2$, the LW background suppresses Pop III star formation in halos with $M_{vir}~<~M_{LW}$ {\it{unless}} $M_{vir} > M_{HI}$.  For a strong enough LW background the redshift at which $M_{LW} = M_{HI}$ is approximately as high or higher than the redshift at which the first halos with $M>M_{HI}$ collapse. As seen in the left panel of Figure~\ref{FIG.Mvir_z_III}, this not only delays the local onset of Pop III star formation to $z~<~15$, but increases the typical halo masses by two orders of magnitude from $\approx~10^5~-~10^6~M_\odot$ to $\approx~10^7~-~10^8~M_\odot$.

The delay of Pop III star formation until $z<15$ seen for our strong LW background is similar to the \cite{Bowmanetal:2018} result placing the start of Pop III star formation at $17~\pm~2$ based on measurements of the width of the HII edge at 78 MHz.

We parameterize whether a given LW background is able to suppress Pop III star formation in all non atomic cooling halos with the relationship between two redshifts, $z_{LW}$ and $z_{HI}$. The first, $z_{LW}$, is the redshift at which the mass threshold determined by the strength of the LW background, $M_{LW}$ (Equation~\ref{EQ.LW}), crosses the atomic cooling threshold. At $z~<~z_{LW}$ the mass threshold which governs which halos are able to form Pop III stars is $M_{HI}$. The second redshift, $z_{HI}$, is the redshift at which the first halos in the simulation exceed the atomic cooling threshold. In our representative box, $z_{HI}~\approx~15$. If $z_{LW}~<~z_{HI}$, then the LW background suppresses Pop III star formation in non-atomic cooling halos, delaying the end of the dark ages. This effect mimics the delay in halo formation and suppression of low mass halos in WDM \citep{Avilia-Reeseetal:2001,Dayaletal:2015}. However, unlike the global effect of WDM, a delay in Pop III star formation due to a strong LW background is a local effect. It will occur in regions which are far enough from early Pop III star formation sites to remain unenriched, yet close enough to be affected by the LW background they generate. Differentiating between the two will require probing the redshifts of Pop III star formation along multiple sight-lines, and in a large range of local environments.\\

\begin{figure}
\centering
\includegraphics[width=\columnwidth]{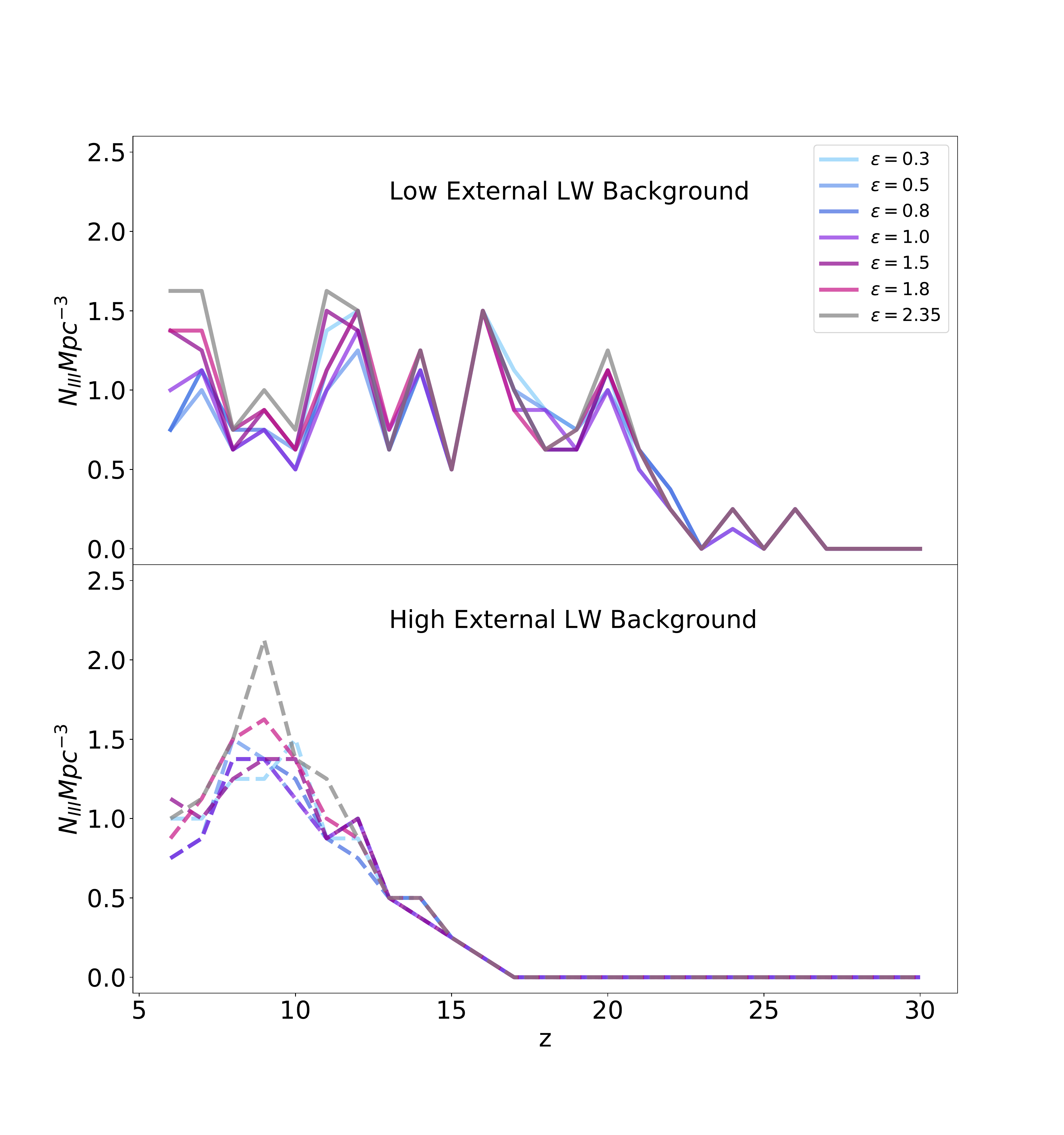}
\caption{Number of Pop III star forming halos per Mpc$^3$ for a range of IMF at a given redshift versus redshift for a weak (upper) and strong (lower) external LW background. Colors of the various IMFs are the same as in Figure~\ref{FIG.III_IMF}. }
\label{FIG.NIII}
\end{figure}

\begin{figure}
\centering
\includegraphics[width=\columnwidth]{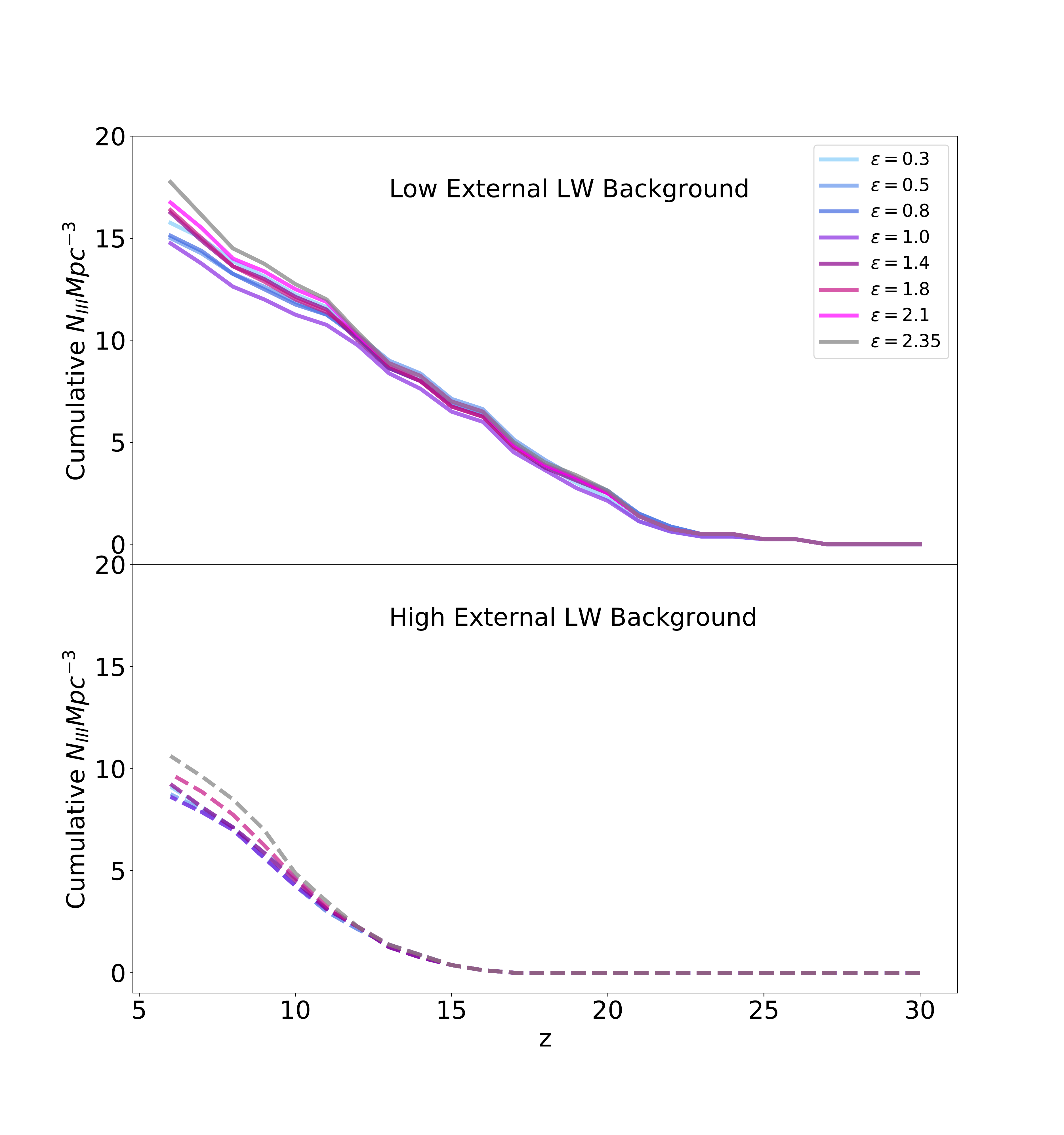}
\caption{Cumulative number of Pop III star forming halos per Mpc$^3$ for a range of IMF at a given redshift versus redshift for a weak (upper) and strong (lower) external LW background. Colors of the various IMFs are the same as in Figure~\ref{FIG.III_IMF}. }
\label{FIG.NIII_cum}
\end{figure}

\subsection{Pop III IMF}
\label{SEC.IMF}

We now look in detail at what variations our model produces in the Pop III IMFs considered in this work. In Figures~\ref{FIG.Mvir_z_III},~\ref{FIG.NIII}, and~\ref{FIG.NIII_cum} we assume our fiducial values for the minimum and maximum masses of Pop III stars of $m_{min}~=~10~M_\odot$ and $m_{max}~=~1000~M_\odot$. 

We first look at variations which are independent of the LW background. While the redshift at which Pop III stars begin to form is independent of the Pop III IMF, Figure~\ref{FIG.Mvir_z_III} shows that there are subtle differences in which halos form Pop III stars for the different IMFs. These differences are highlighted in Figure~\ref{FIG.NIII}, which shows the number density of Pop III star formation halos at $z~<~6$. We find that regardless of the strength of the LW background, the number of Pop III star forming halos at a given redshift decreases as the IMF becomes more top heavy. We now explore the reason behind this trend.

To zeroth order, the number of PISN will increase as the IMF becomes more top heavy. While the energy from CCSNe alone is capable of sending enriched ejecta into the IGM around the lowest mass Pop III star forming halos at $z~>~20$, higher mass Pop III star forming halos at lower redshift require the energy from at least one PISN in order to send enriched ejecta into the surrounding IGM. As the Pop III IMF becomes more top heavy the number of PISN increases and $R_{ejecta}$ becomes larger, decreasing the number of halos at later redshifts which remain pristine.

This scenario also accounts for the larger differences in the number of Pop III star forming halos seen in Figure~\ref{FIG.NIII}. While a larger range of Pop III IMFs produce sufficiently energetic supernova ejecta to enrich nearly halos, only more top heavy IMFs will be able to externally enrich the regions around the more massive Pop III star forming halos forming in the presence of a stronger LW background.

The peaks and valleys seen in Figures~\ref{FIG.NIII} and~\ref{FIG_fkin} for all Pop III IMFs are an effect of LW background in our model. The first Pop III stars form, increasing the LW background, and suppressing Pop III star formation in the lower masses halos, decreasing the number density of Pop III star forming halos and the production of LW photons. Eventually, the original photons will be redshifted out of the LW bands, decreasing the local LW background, and allowing Pop III stars to form in lower mass halos again, increasing their number density.  The location and relative heights of the peaks are dependent on the specific merger trees generated from the N-body simulation. However, not every portion of a JWST NIRCam pointing will be at the same local density. To account for this we assume regions of over densities evolve faster and regions of under-densities evolve slower. To account for this, we can average the number density of Pop III stars over a range of redshifts, which is expected to decrease the amplitude of the spikes. 

\subsection{Fraction of SNe in Mechanical Energy}

In our enrichment model, efficiency with which Pop III supernova can enrich their host halos and the surrounding IGM is direclty dependent on the amount of kinetic energy those supernova inject into the enriched gas. As discussed in S~\ref{SEC.metal}, the total energy generated by Pop III supernova depends on the the number of CCSN and PISN in a given halo. In our model, we assume a fraction of the total supernova energy, $f_{kin}$ goes into kinetic feedback which drives the expansion of the enriched ejecta, and the enrichment of the host halo, and nearby halos. From literature on hydrodynamic simulations \citep{Hopkinsetal:2018,Okuetal:2022,Martizzietal:2015} and other modeling \citep{Kelleretal:2022,Crarinetal:2015,DallaVecchiaetal:2008,KimO:2015}, we have selected three representative values of $f_{kin}~=~0.1~-~1.0$ with a fiducial value of $f_{kin}~=~0.3$.

\begin{figure*}
\centering
\includegraphics[width=0.9\columnwidth]{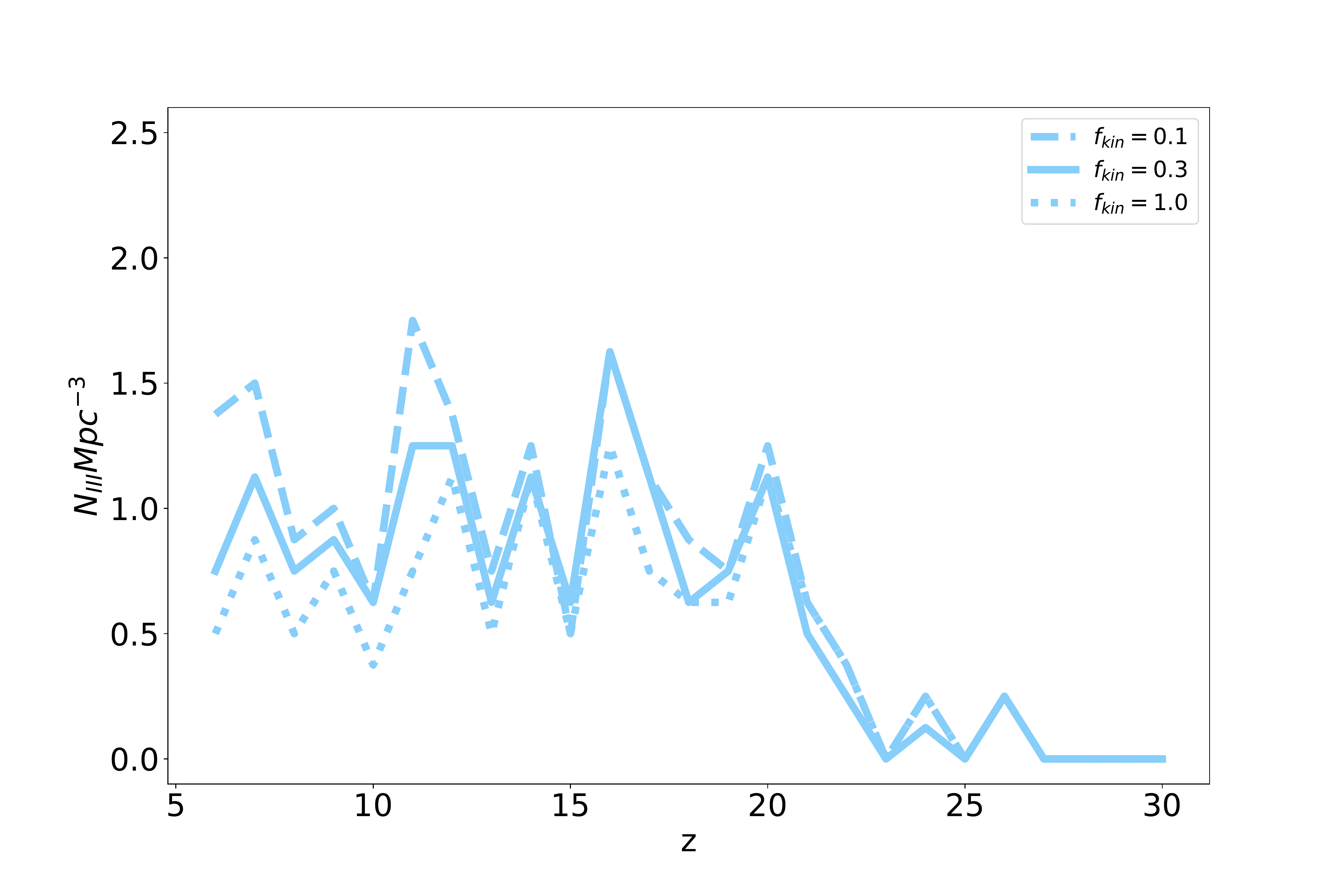}
\includegraphics[width=0.9\columnwidth]{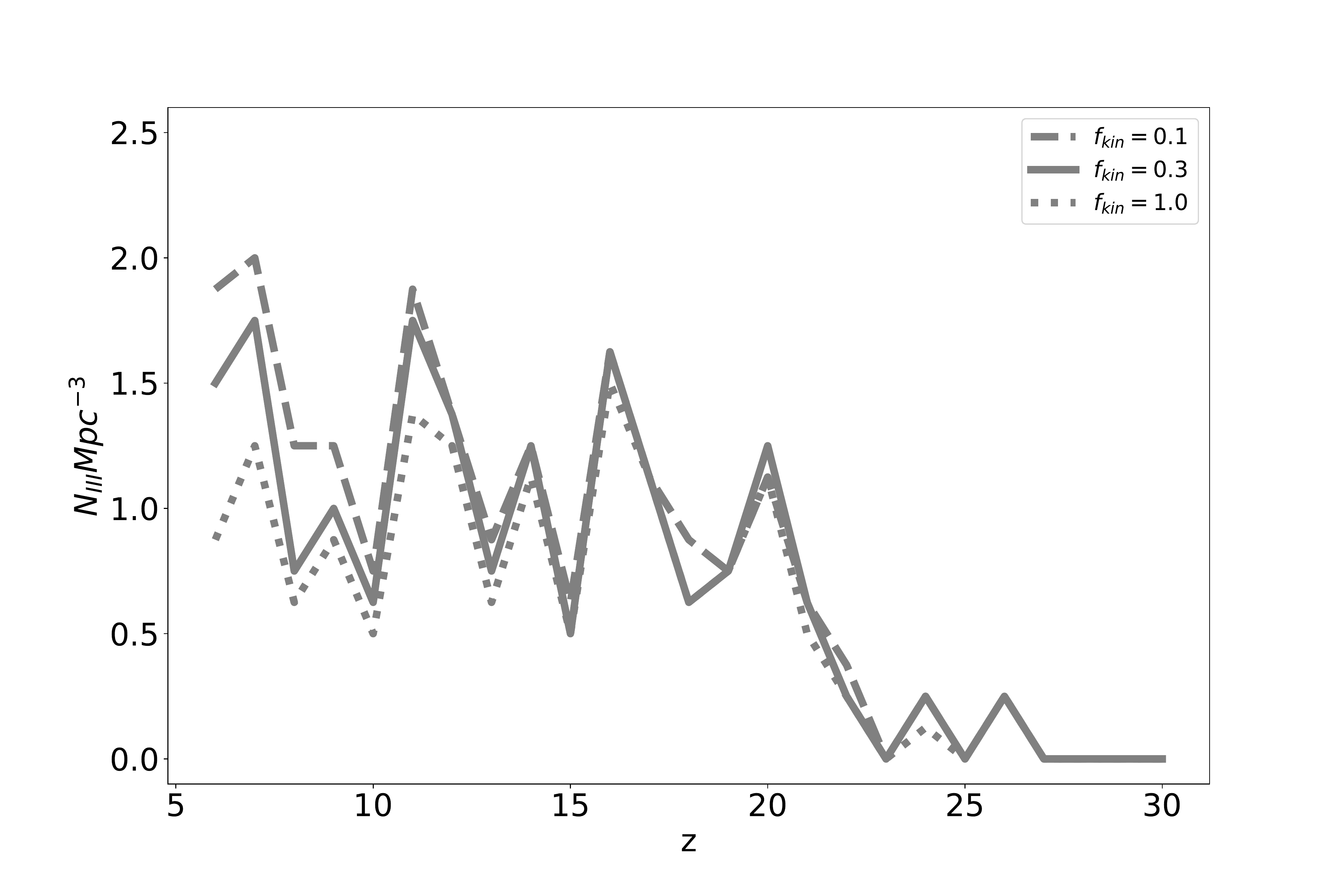}
\caption{Number of Pop III star forming halos per Mpc$^3$ for three possible fractions of supernova energy which go into kinetic feedback ($f_{kin}~=~0.1$ (dashed line), $f_{kin}~=~0.3$ (solid line) and $f_{kin}~=~1.0$ (dotted line)) for $\alpha~=~-0.2$ (left) and $\alpha~=~-2.35$ (right). The range in the number density of Pop III stars versus redshift for the fiducial value of $f_{kin}$ are shown as the colored lines corresponding to the Pop III IMFs in Figure~\ref{FIG.III_IMF}.}
\label{FIG_fkin}
\end{figure*}

Figure~\ref{FIG_fkin} shows the dependence on the number density of Pop III stars on the choice of $f_{kin}$ for two extrema of our Pop III IMFs, $\alpha~=~0.2$ and $\alpha~=~2.35$. As expected, we find the increasing $f_{kin}$ drops the number density of Pop III star forming halos at $z~<~15$. We confirm previous results \citep{Hicksetal:2021,Smithetal:2015,Chenetal:2017,Jaacksetal:2018} which show that the primary driver of the halo and IGM enrichment and the end of the Pop III era is not self enrichment from Pop III star formation in a given halo, but external enrichment of nearby halos by enriched ejecta from Pop III supernova. Therefore, increasing the amount of kinetic energy injected by Pop III supernova will increase the range of the enriched ejecta and decrease the number of halos which retain their primordial composition. 

As seen in Figure~\ref{FIG_fkin}, changing $f_{kin}$ for a given Pop III IMF produces a spread in the number density of Pop III stars which is greater than for various Pop III IMFs for a given $f_{kin}$. However, it is reasonable to assume that $f_{kin}$ is relatively constant in Pop III star forming halos, and the shifts are similar regardless of the chosen Pop III IMF.

\section{Detectability with JWST}
\label{SEC.JWST}

In Figure~\ref{FIG.JWST} we plot the magnitude of Pop III star clusters in the rest frame 1400 A band versus number density of Pop III sources per arcsec$^2$ for our various Lyman Warner backgrounds (colors) and Pop III IMFs (shapes) for three JWST filters, F115W, F150W and F200W. Since we can assume the SED of Pop III stars emits almost all of its light in the UV, each of these filters probes the Pop III population and LW background at a difference redshift. Based on the center wavelengths, F115W probes $z=8.4$, F150W probes $z=10.4$ and F200W probes $z\approx15$.  

We find that our models occupy distinct space in the number density of Pop III sources on the sky versus magnitude space for the F115W, F150W and F200W JWST filters. This shows that when the first Pop III stars are detected their number density and magnitudes will allow us to constrain the LW background at high redshift and/or the Pop III IMF, with a filter dependence on the  effectiveness of the constraint. While too faint for direct detection in even the deepest JWST fields, Pop III star clusters are bright enough to be seen with magnifications typically seen in the Frontier Fields \citep{Zackrissonetal:2015} and in the GLASS ERS observations with JWST \citep{RobertsBorrsanietal:2022}.

\begin{figure*}[ht]
\includegraphics[width=\textwidth]{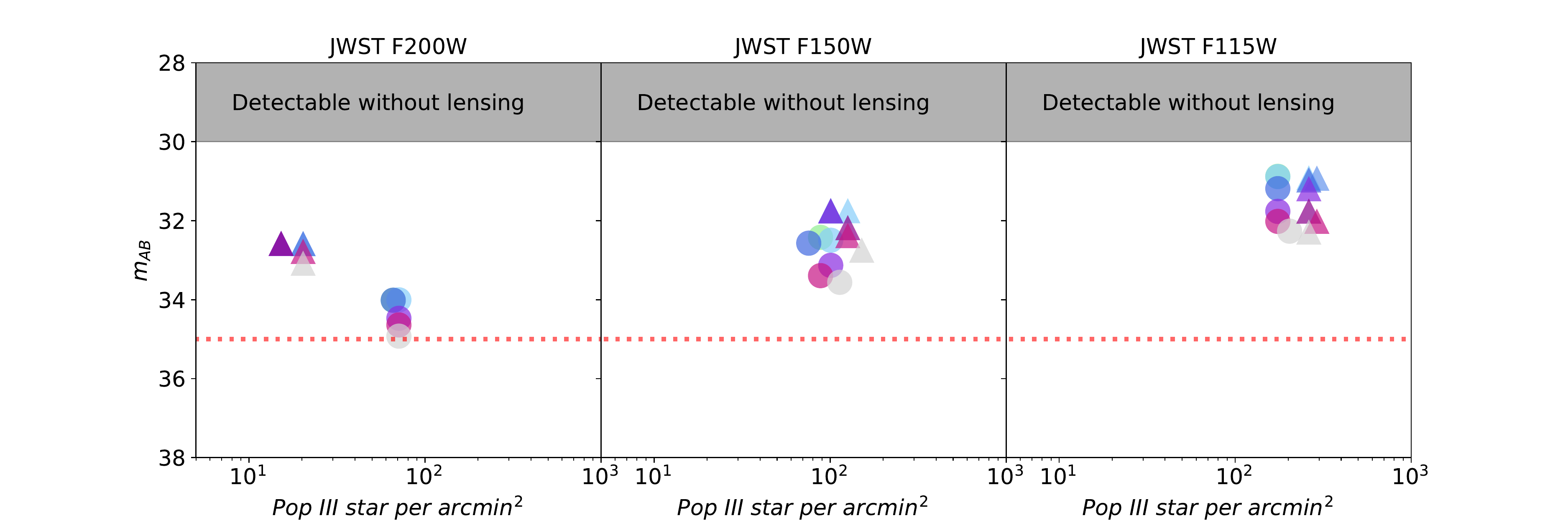}
\caption{Apparent AB magnitude versus number of Pop III star forming halos per arcmin$^2$ assuming a maximum Pop III stellar mass of $1000~M_\odot$ for our range of Pop III IMF slopes (colors from Figure~\ref{FIG.III_IMF}) and a weak (circles) an strong (triangles) external LW background. The shaded region shows magnitudes where a Pop III star cluster would be detectable by a deep NIRCAM pointing (dark grey), the maximum magnification from gravitational lensing typical of the Frontier Fields is the dotted red line. Pop III star clusters below this line are only detectable via rare caustic crossing events. We show the relation for three NIRCAM filters corresponding to a $\Delta z~\approx~1$ centered on $z~=~15.4$ (left), $z~=~11.3$ (center) and $z~=~8.4$ (right).}
\label{FIG.JWST}
\end{figure*}

\begin{figure*}[ht]
\includegraphics[width=\textwidth]{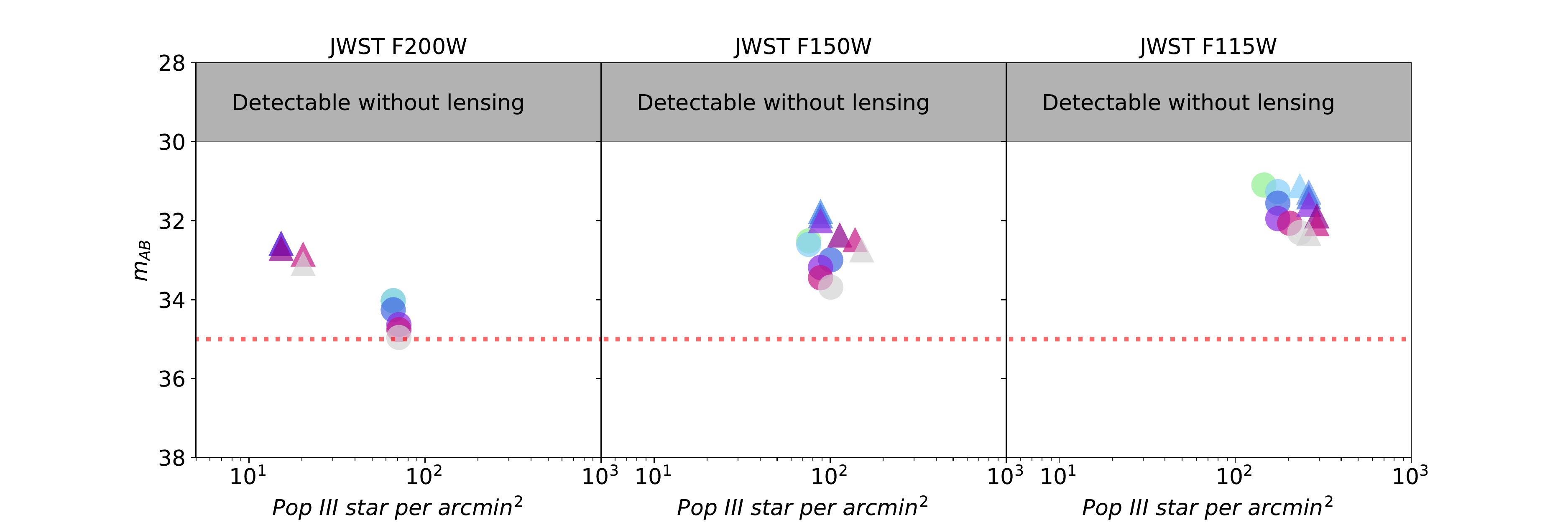}
\caption{Apparent AB magnitude versus number of Pop III star forming halos per arcmin,$^2$ assuming a maximum Pop III stellar mass of $300~M_\odot$ for our range of Pop III IMF slopes (colors from Figure~\ref{FIG.III_IMF}) and a weak (circles) and a strong (triangles) external LW background. The shaded region shows magnitudes where a Pop III star cluster would be detectable by a deep NIRCAM pointing (dark grey), the maximum magnification from gravitational lensing typical of the Frontier Fields is the dotted red line. Pop III star clusters below this line are only detectable via rare caustic crossing events. We show the relation for three NIRCAM filters corresponding to a $\Delta z~\approx~1$ centered on $z~=~15.4$ (left), $z~=~11.3$ (center) and $z~=~8.4$ (right).}
\label{FIG.JWST}
\end{figure*}

We find, regardless of the LW background and Pop III IMF, there are an order of 100-1000 Pop III star forming halos per rNIRCam pointing in all three filters. In F115W, there are $\sim1000$ Pop III star forming halos per NIRCam pointing for the low LW background and $\sim1500$ Pop III star forming halos per NIRCam pointing for our high LW background. 

In F150W we find $\sim500$ Pop III star forming halos per NIRCam pointing for both LW background strengths, though in general the high LW background produces a slightly higher number density of more luminous star forming halos in the filter. This is expected given that at $z\sim11$ the delay in Pop III start formation for the stronger LW background means Pop III stars will be forming in higher mass pristine halos in greater number densities than for the lower LW background where many of these atomic cooling halos are enriched.

It is in F200W, probing $z\sim15$, where the strength of the LW background produces an order of magnitude differences in the number of Pop III star forming halo per NIRCam pointing. While the low LW background produces $\sim350$ Pop III star forming halos per NIRCam pointing, the suppression of star formation in low mass halos by the high LW background means that for the stronger LW background there are only $\sim50$ Pop III star forming halos per NIRCam pointing.

These results underscore the importance of observations across multiple filters in order to disentangle environmental effects like the LW background from the Pop III IMF in $m_{AB}$-number density space. A weaker LW background will produce a steady increase in both number density and magnitude from F200W to F115W. In contrast, the suppression of Pop III star formation in low mass halos by a strong LW background produces a lesser increase in magnitude but an increase in number density of a factor of 20 across the same filters. How these trends translate to observations of magnified Pop III stars in the Frontier Fields will be the subject of an upcoming paper.

For all JWST filters, we find more top heavy IMFs produce more luminous Pop III star forming halos irrespective of the local LW background and the maximum mass of the Pop III IMF. This is a consequence of the consideration of both the Pop III star formation efficiency and the fragmentation limits of primordial gas in our model. As seen in Figure~\ref{FIG.SFE}, the defacto Pop III star formation efficiency of Pop III star formation increases as the IMF becomes more top heavy. For the more bottom heavy IMFs, the total mass in Pop III stars is limited by the fragmentation of primordial gas, limiting their luminosities.

\section{Conclusions}

We have presented the initial results from a novel model of Pop III star formation which accounts for the limitations of the fragmentation of primordial gas while treating the Pop III IMF and maximum mass as a truly free parameters. While relatively simple, our model is able to roughly reproduce the Pop III star formation rates seen in \cite{SkinnerW:2020} for an equivalent IMF. In this work, we use our model to explore Pop III star formation for a range of IMFs.

A critical component of our work is the inclusion of a model for the fragmentation of primordial gas based on simple assumptions for Jeans collapse in $>~10^7~M_\odot$ halos. We find that the expected lower fragmentation of primordial gas limits the number of Pop III stars which can form, especially for the more bottom heavy IMFs. Our results suggest that the shape of the Pop III IMF may be driven by the efficiency of the fragmentation of primordial gas. In halos where higher angular momentum and turbulence have driven additional fragmentation, the Pop III IMF may form Pop III stars of lower masses. The full implications of a potentially non-universal Pop III IMF and a more detailed study of how the efficiency of primordial gas fragmentation compares to our simple model for various halo mass and redshifts are both subjects for future study.

If direct detection is possible, the number density of Pop III stars at a given redshift in one NIRCam pointing is greater than one, and greater than 10 for certain combination of LW background and Pop III IMFs.

We find that, while the number density of Pop III stars is relatively independent of the mass distribution of Pop III stars, the magnitude of Pop III star clusters is $2~-~3$ magnitudes higher for top heavy Pop III IMFs than for their more bottom heavy counterparts. In addition, this result is relatively independent of the assumption for the maximum mass of Pop III stars.

While we find that all Pop III stars are too faint for direct detection by JWST, they are bright enough to be detected with magnification possible in the Frontier Fields for all combinations of Pop III IMF and LW background explored in this study.  A quantification of how the dependence of the magnitudes of Pop III clusters on the IMF will affect the detectability of the first stars in upcoming JWST observations of lensing clusters will be the subject of a future study.\\

\noindent The authors acknowledge the University of Maryland supercomputing resources (http://hpcc.umd.edu) made available for conducting the research reported in this paper.

\bibliography{references}{}
\bibliographystyle{aasjournal}

\end{document}